\documentclass[hidelinks,lettersize,journal]{IEEEtran}

\usepackage{amsmath,amsfonts}
\usepackage{array}
\usepackage{stfloats}
\usepackage{graphicx}
\usepackage{cite}
\usepackage{balance}

\AtBeginDocument{
  \providecommand\BibTeX{{
    \normalfont B\kern-0.5em{\scshape i\kern-0.25em b}\kern-0.8em\TeX}}}

\usepackage{algorithm}
\usepackage{algpseudocode, tabularx}
\usepackage{amsmath}
\usepackage{adjustbox}
\usepackage{subcaption}
\usepackage{colortbl} 
\usepackage{multirow}
\usepackage{hhline}
\usepackage{enumitem} 
\usepackage[innerleftmargin=5pt, innerrightmargin=5pt, middlelinewidth=0.8pt]{mdframed} 

\newcolumntype{C}{>{\centering\arraybackslash}X} 

\usepackage[hidelinks]{hyperref}
\usepackage{tikz}

\makeatletter
\newcommand{\multiline}[1]{%
  \begin{tabularx}{\dimexpr\linewidth-\ALG@thistlm}[t]{@{}X@{}}
    #1
  \end{tabularx}
}
\makeatother

\begin{document}

\title{MTDSense: AI-Based Fingerprinting of Moving Target Defense Techniques in Software-Defined Networking}

\author{
    \href{https://orcid.org/0000-0002-7234-9127}{Tina~Moghaddam},
    \href{https://orcid.org/0000-0002-1404-4560}{Guowei~Yang},
    \href{https://orcid.org/0000-0002-3855-3378}{Chandra~Thapa},
    \href{https://orcid.org/0000-0001-6353-8359}{Seyit~Camtepe},
    and~\href{https://orcid.org/0000-0003-2605-187X}{Dan~Dongseong~Kim}

    \thanks{ T. Moghaddam is with the School of Electrical
    Engineering and Computer Science, University of Queensland, St Lucia, QLD 4072, Australia, and also with Data61, Commonwealth
    Scientific and Industrial Research Organization, Sydney, NSW 2122, Australia (e-mail: \href{mailto:t.moghaddam@uq.edu.au}{t.moghaddam@uq.edu.au}).}
    \thanks{C. Thapa and S. Camtepe are with Data61, Commonwealth
    Scientific and Industrial Research Organization, Sydney, NSW 2122, Australia
    (e-mail: \href{mailto:chandra.thapa@data61.csiro.au}{chandra.thapa@data61.csiro.au}; \href{mailto:seyit.camtepe@data61.csiro.au}{seyit.camtepe@data61.csiro.au}).}
    \thanks{G. Yang and D.D. Kim are with the School of Electrical
    Engineering and Computer Science, University of Queensland, St Lucia, QLD 4072, Australia (e-mail: \href{mailto:guowei.yang@uq.edu.au}{guowei.yang@uq.edu.au}; \href{mailto:dan.kim@uq.edu.au}{dan.kim@uq.edu.au}).}
    \thanks{\textit{(Corresponding author: T. Moghaddam)}}
}
\markboth{}%
{}

\maketitle
\begin{abstract}
Moving target defenses (MTD) are proactive security techniques that enhance network security by confusing the attacker and limiting their attack window. MTDs have been shown to have significant benefits when evaluated against traditional network attacks, most of which are automated and untargeted. However, little has been done to address an attacker who is aware the network uses an MTD.  In this work, we propose a novel approach named \mbox{MTDSense}, which can determine when the MTD has been triggered using the footprints  the MTD operation leaves in the network traffic. MTDSense uses unsupervised clustering to identify traffic following an MTD trigger and extract the MTD interval. An attacker can use this information to maximize their attack window and tailor their attacks, which has been shown to significantly reduce the effectiveness of MTD. Through analyzing the attacker's approach, we propose and evaluate two new MTD update algorithms that aim to reduce the information leaked into the network by the MTD.  We present an extensive experimental evaluation by creating, to our knowledge, the first dataset of the operation of an IP-shuffling MTD in a software-defined network. Our work reveals that despite previous results showing the effectiveness of MTD as a defense, traditional implementations of MTD are highly susceptible to a targeted attacker.
\end{abstract}


\section{Introduction}
\IEEEPARstart{A}{s} the world grows increasingly digital and more of the everyday workings of life and business come to rely on computer networks, securing network infrastructure is an ever-present problem. In this landscape, traditional networks resemble sitting ducks, as their static nature gives attackers an advantage \cite{sengupta_survey_2020}. More modern virtualized networks still underutilize their potential to respond to adaptive attacks by adopting traditional modes of operation despite their flexibility.  
Moving Target Defense (MTD) is a class of defense strategies that allow the defender to be more proactive \cite{cho_toward_2020}. In adopting MTD, the defender changes some properties of the attack surface to hinder the attacker. Proposed MTDs have changed the network hosts' IP addresses, transmission routes, and available applications to invalidate the attacker's understanding of the network, reduce their windows of opportunity,  and confuse them in their approach. Recently, MTDs are also increasingly being offered as consumer security solutions through companies such as SCIT Labs \cite{scit_labs_scit_2024} and NexiTech \cite{nexitech_moving_2023}. As the defense and commercial usage of MTD increases, evaluating their true security against sophisticated attackers is increasingly important.

Since their introduction, MTDs have been shown to be effective in reducing the attacker's chances of success \cite{cho_toward_2020, sengupta_survey_2020}. However, most evaluations assume that the attacker does not consider the MTD itself or does not even know it is being used as a defense mechanism.  Previous work has mostly evaluated MTD against simple automated attackers, such as scanning worms and DDoS attacks  \cite{cho_toward_2020, cai_moving_2016}. Very few works consider AI-powered attackers who can adjust their behavior given the MTD. Among these, skilled human attackers have been considered \cite{jafarian_multi-dimensional_2016}, as well as deanonymizing devices employing MAC address randomization \cite{vanhoef_why_2016, matte_defeating_2016}. To the best of our knowledge, no work directly targets network MTD itself.

This paper addresses the gap by exploring how an attacker could overcome the MTD by targeting it directly in reconnaissance. Specifically, we look in detail into how the attacker can determine when an IP shuffling MTD is triggered in a software-defined network (SDN). Machine learning techniques are making it much simpler to find meaningful information among large volumes of noisy data at near-real-time speeds. We leverage this in analyzing passively collected network traffic between legitimate clients and servers. We propose a novel approach called \mbox{MTDSense}, which uses unsupervised clustering techniques to determine the absolute MTD trigger time and the duration between triggers, which we call the MTD interval. With this information, an attacker can plan the timing of their attacks, which has been shown to improve the attack success rate across all phases of a cyber attack by 40\% \cite{moghaddam_practical_2022}. Our work reveals that current MTD mechanisms are susceptible to attackers who target them during reconnaissance, even though there has been considerable work to show that they are effective against common known attacks.  

Next, we investigate the fingerprint the MTD leaves in the network that allows the MTD trigger to be detectable using machine learning. By finding the unique timing features that reveal the operation of the MTD, we propose alternative update mechanisms to reduce the MTD fingerprint. Typical mechanisms for IP shuffling MTD are modeled on common routing protocols of the internet and do not consider information leakage. Our novel update mechanisms deliberately time the update operations to minimize the information leak and make it harder for MTDSense to detect their operation. We experimentally evaluate MTDSense with respect to network and traffic size and MTD update mechanism using a real SDN testbed as well as supporting simulations.

The main contributions of this work are summarized as follows: 
\begin{itemize}[noitemsep,topsep=0pt]
    \item We demonstrate that an Artificial intelligence (AI)-capable attacker can determine the timing of an IP shuffling MTD.  To the best of our knowledge, this is the first work where network reconnaissance targets the MTD directly. 
    \item We propose, implement, and evaluate two novel MTD update algorithms with the aim of reducing the MTD fingerprint. The susceptibility of these algorithms against the attack and the respective trade-offs are analyzed.
    \item We collect a dataset from an  SDN using an IP shuffling MTD for defense covering both a simulated and testbed environment. The dataset varies a series of parameters of network and traffic size and comprises over 150 days of traffic data. 
\end{itemize}

\section{MTD in Software Defined Networking} 
\label{Section:MTD_in_SDN}

While MTD techniques have been studied in other systems, software-defined networking (SDN) introduced a new set of MTD techniques for large networks and accelerated research with the development of the OpenFlow protocol standard \cite{cho_toward_2020}. SDN decouples the control of the network from the physical devices. 
Decision making in SDNs is centralized to the SDN controller. In turn, SDN-enabled switches store network traffic rules in their flow tables, which can be programmed by the controller.  
This makes it possible to change system configurations on the fly and more frequently than in traditional decentralized systems, making implementing MTDs in SDNs much easier in comparison. As a result, a large portion of MTDs are implemented in SDNs \cite{sengupta_survey_2020}. 

Network level and specifically network address randomization MTD techniques are amongst the most popular and most researched MTD techniques \cite{sengupta_survey_2020, cho_toward_2020}.  
In IP shuffling MTD, also known as host address mutation MTD, network hosts are assigned a temporary IP address that is valid for only a certain period of time. These techniques are designed to counter network reconnaissance. While an attacker scans a network address range and proceeds with their attack based on this information, the MTD is triggered, and the IP addresses and any information about the hosts with a particular IP that were gathered by the attacker are no longer valid. Thus, the attacker must spend more resources and has a valid view of the network for only a limited time.  In this work, we use an IP shuffling MTD based on Flexible Random Virtual IP Multiplexing (FRVM), where each port on each host is assigned a virtual IP address (vIP) \cite{sharma_frvm_2018}. Evaluations of this technique have shown it to be effective at reducing the attacker's success in network reconnaissance through modeling \cite{sharma_frvm_2018}, emulations \cite{dishington_security_2019}, and in a realistic testbed \cite{kim_performance_2022}.  

Each network host has a real IP address (rIP). The SDN controller assigns each host with virtual IP addresses and maintains a mapping between the rIPs and vIPs. When the MTD is triggered, new vIP addresses are assigned to the hosts. The controller adjusts the flow rules in the SDN-enabled switches to forward packets according to the mapped vIPs. The replacement of the rIP and vIP in the IP packets is done at the network edge, so the IP mutation is transparent to both end-user devices and network hosts. Legitimate clients can find the new vIPs through an authenticated DNS service. 
Any hosts scanning the network will not be aware that their previously discovered hosts are invalid when the IP shuffling has occurred. 

The MTD triggering can be time-based, happening at regular intervals (which we call the MTD interval $T$),  event-based,  happening after some network event, such as an IDS alert,  or a hybrid of the two. A shorter MTD interval will provide greater security against scanning attacks but also incur a greater performance cost \cite{kim_performance_2022}.

\section{MTD Architecture and Implementation} \label{Section:archi}

\begin{figure}
\centering
    \subfloat[]{\includegraphics[width=0.45\textwidth]{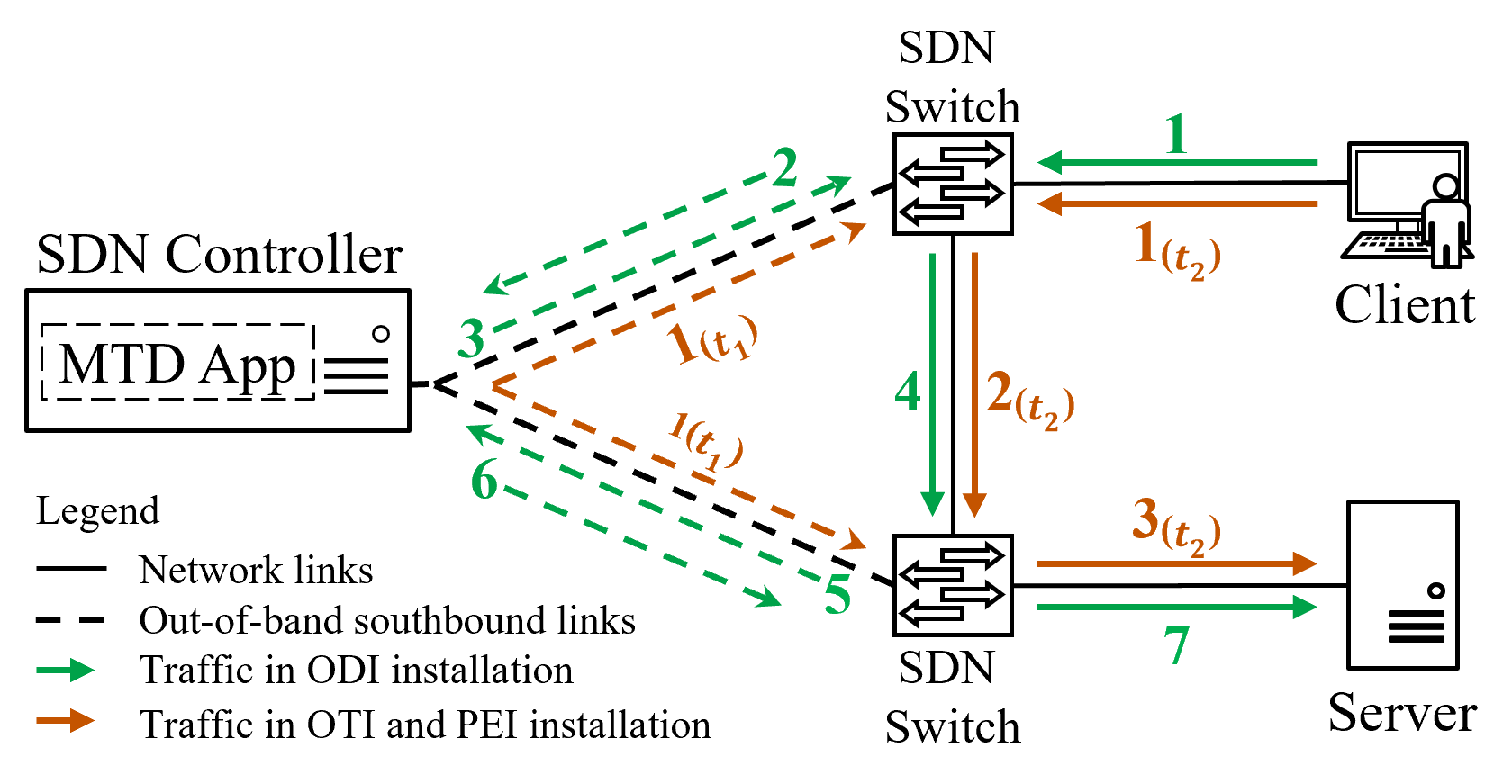}
    \label{fig:architecture:nw}}
\hfil
    \subfloat[]{\includegraphics[width=0.35\textwidth]{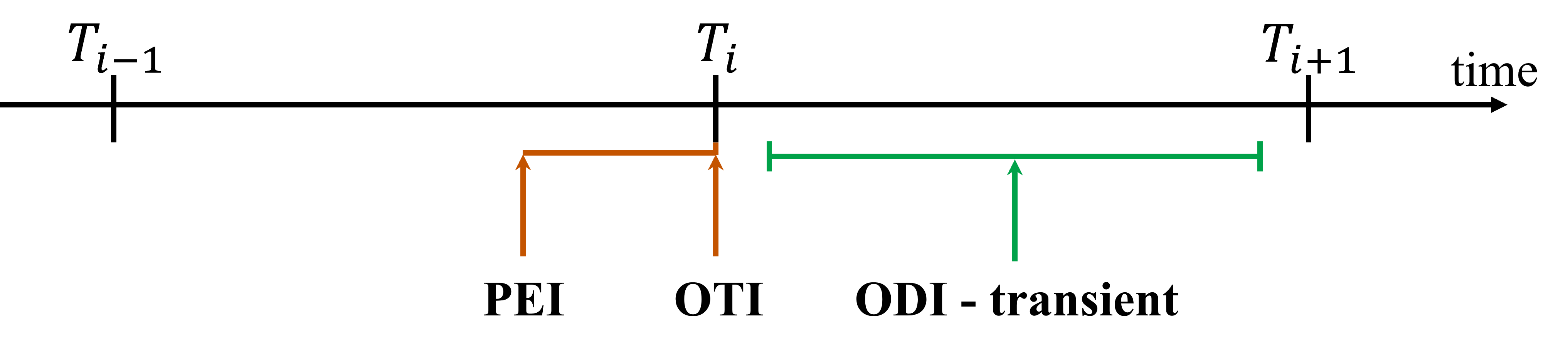}
    \label{fig:architecture_timing}}
\caption{\textbf{(a)} Steps of MTD update at $T$ based on the update mechanism. Lines above the network connections (green) show the steps for on-demand installation (ODI), and the lines below the network connections (orange) show the steps for on-time installation (OTI) and PEI MTD, which happen at two distinct times $t_1$ and $t_2$. \textbf{(b)} Timing of when the MTD update is made in the SDN switches for update $T_i$. In the ODI scheme, updates are made when a client connection is requested, making it transient. In PEI, updates are installed in advance but do not come into effect until $T_i$.}
\label{fig:architecture}
\end{figure}

The MTD design defines the broad way the MTD should operate: when it should trigger, what should change in the system, and how it should change. The implementation details of how the MTD is triggered influence the fingerprint the operation leaves in the network traffic, which can then be exploited by an attacker. We call the sequence of events that are enacted at MTD trigger time the `update mechanism.' Below, we categorize and explore three possible update mechanisms and their characteristics.

\subsection{On-Demand Installation}
In on-demand installation (ODI), the controller removes any MTD-related rules in the switches at MTD trigger time. The switches always retain a rule to forward any packets that do not match any other rules to the controller. As such, when a new packet arrives at a switch with no corresponding rule, it is forwarded to the controller, who decides what to do. The MTD application in the controller computes what IP substitutions need to be made in the packet and installs a corresponding rule in the switches. Thus, MTD rules are installed on the switches when there is relevant traffic. The algorithm for this update mechanism is shown in Algorithm \ref{alg:od_install}. The previous rules are deleted after the MTD trigger interval $T$ has passed, and the rules are installed in each switch after a packet arrives at some other time.  This typical mode of operation is described in \cite{dishington_security_2019}.

\begin{algorithm}
    \caption{MTD trigger algorithm for on-demand installation}
\begin{algorithmic}
    \State SrcIP: the source IP address for traffic
    \State DstIP: the destination IP address for traffic
    \Require $\Delta t \geq T$
    \State $switches \gets $ deleteRules
    \State updateDNS
    \Require $packet$ at $switch$   
    \If{Src or Dst is in Network} 
        \If{Src is in Network}
            \State $switch \gets$ newRule(\multiline{forward $rIP$ to $vIP$\\ for $DstIP$)}
        \ElsIf{Dst is in Network}
            \State $switch \gets$ newRule(\multiline{forward $vIP$ to $rIP$\\ for $SrcIP$)}
        \EndIf
        \State \textbf{Await}  confirmation from $switch$
        \State \textbf{Forward} $packet$ along path to next $switch$
    \Else 
        \State \textbf{Drop} $packet$ 
        \Statex \Comment{We can check for malicious or irrelevant traffic.}
    \EndIf
\end{algorithmic}
\label{alg:od_install}
\end{algorithm}

This method mimics the common mechanisms that are used for maintaining routing tables and forwarding traffic on the internet, similar to updating addresses in layer-2 switches. The majority of the logic is independent for each switch and path, and the MTD operation interfaces only minimally with the normal flow of traffic. As a result, this method is very flexible and has minimal overhead. The flexibility means that, like in traditional networks, switches and hosts can be added at any time, and there is minimal impact on load balancing. 

Some aspects of the operation of the MTD are also simpler in this scheme. Legitimate IP addresses do not need to be tracked in advance and instead can be checked at arrival time. Inserting rules has a performance cost, and once rules are added, the performance of a switch is inversely related to the size of its flow table. In this scheme, rules are only installed when needed, so the impact on performance is minimized.   

From a security perspective, on-demand installation allows the controller to try and distinguish between a legitimate client and an attacker at packet arrival time before installing the rules that allow access to network hosts. There are no additional components or databases needed to implement this method, and the details of the implementation are clear. Since the changes to the switches are dependent on the incoming traffic, the operations are naturally staggered in time, giving better resource utilization. However, this also means that there is no one singular MTD trigger time; rather, it is a transient period of time after $T$ when the rules are most likely being installed. Based on our analysis of ODI with MTDSense, we propose on-time and pre-emptive installation, which we describe next.

\subsection{On-Time Installation}
In on-time installation (OTI), the switches are emptied and repopulated at the MTD trigger time $T_i$. As all of the MTD operations happen at the start of an interval, this is the only time that any fingerprints can be left in the network traffic. Once the installation period has passed, there will be no further MTD operations until the next trigger interval starts. The algorithm for this is shown in Algorithm \ref{alg:p_install}. 
Once again, the rule to forward nonmatching packets to the controller is retained. In this scheme, any traffic that matches no rules will be forwarded to the controller, where it can be managed according to the firewall rules. In this case, the traffic will be dropped if the client cannot authenticate, or it will be forwarded through the network by the controller, similar to the ODI scheme. 

\begin{algorithm}
    \caption{MTD trigger algorithm for on-time installation}\label{alg:p_install}
\begin{algorithmic}
    \Require $\Delta t \geq T$
    \State $switches \gets$ deleteRules 
    \ForAll{$switch \in switches$} 
        \ForAll{$client \in ExpectedClients$} 
            \ForAll{$rIP \in rIPs$}
                \State $switch \gets$ newRule(\multiline{forward $rIP$ to $vIP$\\ for $DstIp$)}
                \State $switch \gets$ newRule(\multiline{forward $vIP$ to $rIP$\\ for $SrcIp$)}
            \EndFor
        \EndFor
    \EndFor
    \State updateDNS    
\end{algorithmic}
\end{algorithm}

This method ensures all new rules are installed at the switches some short time after the MTD trigger time. However, it is less flexible. Since the clients are not being discovered and authenticated at connection time, the controller must maintain a list of expected clients (both previously connected clients and those that could be expected to connect in this MTD interval).   The controller needs to decide in advance which clients to allow and precalculate the paths through the network.  The operations at MTD trigger time will also take longer as many rules need to be installed at once. In the case of event-based MTD, using this mechanism also means there will be a longer delay when triggering the MTD after an event.  

The lack of flexibility from this method means some downsides need to be mitigated. Since all clients are added at the same time this hinders the controllers ability to apply up to date load balancing, instead any load differences must be predicted in advance. Additionally, there may be a longer delay in making changes if the optimal path changes due to network events. These issues can be addressed at the next MTD trigger instead of at any time, and the MTD intervals may be short enough for this to be acceptable. Finally, in a large network, precalculating the paths can have a large storage overhead and require a lot of computing time. 

\subsection{Pre-Emptive  Installation}
The goal of the pre-emptive installation (PEI) method is to have the new rules come into effect at trigger time $T_i$, with any operations finalized by this time. Like in the OTI scheme, this again requires the controller to predetermine and authenticate clients. In this scheme, sometime before $T_i$, the controller installs the rules for the next MTD interval into the switches. The rules are configured to come into effect at the next MTD interval start time and expire at the interval end time. The algorithm for this scheme is shown in  Algorithm \ref{alg:pe_install}.

\begin{algorithm}
    \caption{MTD trigger algorithm for pre-emptive installation}\label{alg:pe_install}
\begin{algorithmic}
    \Require $\Delta t < (T-\epsilon)$
    \ForAll{$switch \in switches$} 
        \ForAll{$client \in ExpectedClients$}
            \ForAll{$rIP \in rIPs$}
                
                \State $switch \gets$ newRule(\multiline{forward $rIP$ to $vIP$ \\ for $DstIp$, expire after $T$)}
                
                \State $switch \gets$ newRule(\multiline{forward $vIP$ to $rIP$ \\ for $SrcIp$, expire after $T$) 
                }
                \Statex \Comment{Rules are added at the bottom of the \\ \hfill flow table.}
            \EndFor
        \EndFor
    \EndFor
    \Require $\Delta t \geq T$
    \State updateDNS  
\end{algorithmic}
\end{algorithm}

In this way, the overhead of communication between the controller and the switches is taken on before the trigger time, so the actual delay for the change between the two sets of rules should be reduced. This requires that the switch allows the controller to install new rules without needing to reload or otherwise interfere with ongoing traffic. This method suffers from the same reduced flexibility as on-time installation. 

In addition, there are security implications for PEI and, to some extent, OTI implementations. By preempting the MTD trigger and installing all the possibly required rules, the information about the configurations for the next MTD interval is available more completely and for a longer time. The full list of authenticated clients is also stored in two places and is now also being kept on the switches at all times. Keeping in mind that SDN switches can also be compromised \cite{lee_delta_2017}, this increases risk. Further, connecting devices cannot be re-assessed for access at connection time. These issues are bounded by the limited MTD interval, however additional access control mechanisms need to be implemented to manage dynamic changes during the interval.

Figure \ref{fig:architecture} shows a simplified model of the SDN environment, with the timings and behavior for the three update mechanisms. In the ODI case, all communication from the controller is made whenever traffic arrives at a switch. In the OTI and PEI cases, there is an initial time $t_1$ when all rules are installed and a later time $t_2$ when traffic arrives at the network. The difference in timing of the update with respect to the MTD interval is shown in Figure \ref{fig:architecture_timing}. 

\subsection{Implementation and Data}

\subsubsection{Environment and Component Behavior}
\begin{figure}[t]
    \centering
    \includegraphics[width=0.4\textwidth]{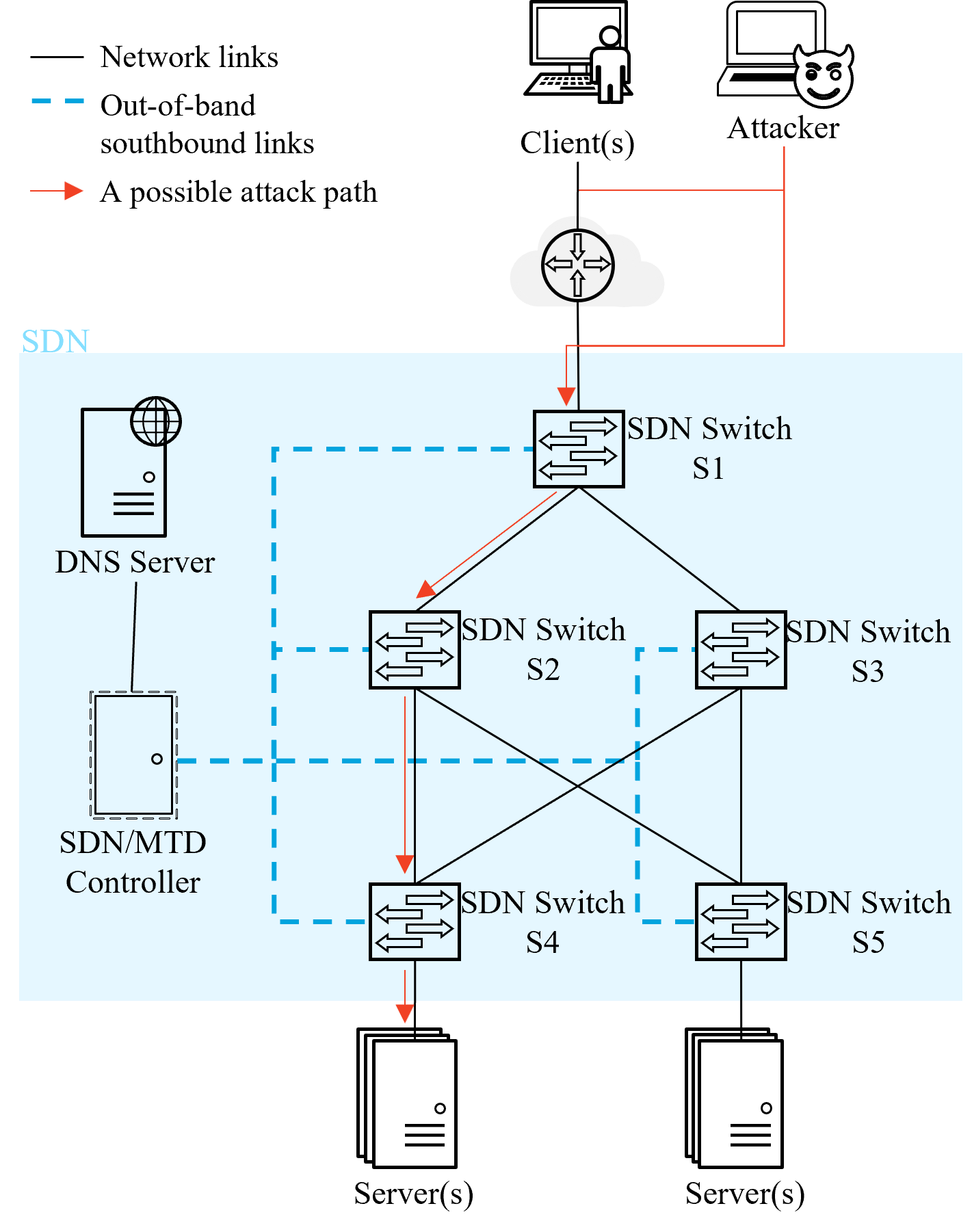}
    \caption{Our experimental environment.}
    \label{fig:mininet_testbed}
\end{figure}

Data was collected in a small-scale SDN testbed, which offers the necessary dynamism and flexibility to effectively evaluate the performance and behavior of the MTD and the attacker. Because testbed experiments must run in real time and are therefore limited, we supplement the data with simulations using Mininet \cite{mininet_2022}. 
Mininet was carefully configured to simulate the testbed environment. These cover intermediate parameter values and allow for faster prototyping during development. 
The general structure of the network is shown in Figure \ref{fig:mininet_testbed}. It includes the following components:
\begin{itemize}
    \item \textbf{The controller}, which is connected to the switches through separate out-of-band links and implements the MTD.  We use a Open Network Operating System (ONOS) 2.4.0 controller \cite{onos} and OpenFlow 1.3.

\item \textbf{A DNS server}, which provides service to legitimate clients.

\item \textbf{A router}, which divides the SDN network from the user network and acts to simulate a connection over the internet.   

\item \textbf{SDN-enabled switches} that are programmed by the controller.  In the SDN-testbed, these are HP Aruba 2920F and 2930M network switches.

\item \textbf{Servers} running an Apache webserver \cite{apache}. 

\item \textbf{Clients} who periodically connect to the servers and request webpages of various sizes.  

\item \textbf{The attacker} which sits outside the SDN network and is connected to the same switch as the client. 

\end{itemize}

\subsubsection{Threat Model}
The assumption of compromise is that the attacker needs to be in the same subnet as a legitimate client and be able to sniff the client's traffic. The attacker could be located anywhere in the network and could equally compromise the client, a switch between the client and the network, or any other elements on the path, but it does not need to (though it could) compromise a component inside the SDN network. The variety of  choices means that it is more likely the attacker could achieve access to at least one of these components, and vulnerabilities are commonly found in various components of networks, including in the SDN components themselves \cite{yoon_flow_2017}. The attacker passively sniffs the legitimate client's traffic, and since it is only sniffing passively and is outside the protected network, it cannot be detected without extensive active scanning. We further relax the requirements by showing that the traffic need not necessarily come from one client but can be aggregated from multiple clients of the network.
The data collection can be located anywhere, does not need to come from a singular source,  does not involve active disruption to the network and so does not need to evade detection by the SDN network.

\subsubsection{Parameters and Data Collection} \label{Section:implementation:params}
Table~\ref{fig:parameters} shows the different experimental parameters that were considered when taking data and the ranges of values that are considered for each of the variables. What follows is an explanation of each parameter and the reasoning for the chosen values. 

The attacker observation window $O_{\textup{win}}$ is the duration of time the attacker sniffs network packets before trying to determine the MTD properties. This is varied between $3 \times the\  mtd\  interval$, which is guaranteed to capture 2 MTD triggers, and so is considered the lowest possible duration for calculating the interval, and 10 hours. These values are used to evaluate how long an attacker needs to observe a network to be able to deduce MTD properties with our method. 
The MTD interval $T$ is the duration between two MTD triggers. The range here was chosen based on the analyses by Connell \textit{et al.} 
\cite{connell_performance_2021},  Mendonça  \textit{et al.} \cite{mendonca_performability_2020} and Kim  \textit{et al.} \cite{kim_performance_2022}, who have analyzed the performance and security trade-off of the MTD interval which varies with respect to the expected attack length. Based on their analyses, we find an interval of 60-300s to be appropriate for our case and extend our window slightly in both directions to cover other cases.

Requests by the client are a Poisson process defined by the mean  $\lambda$.  The highest value for  $\lambda$ was chosen based on benchmarking of the testbed. Due to the limitation of the simulated network, higher values of lambda lead to occasional network dropouts. The slowest client request interarrival time is half of the MTD interval, which is very slow considering the number of requests fielded by a typical server on the internet, but still ensures there is at least one request in each MTD interval, which is required for triggering in the ODI case. The lower bound of 10 requests per second is typical in the literature  \cite{kim_performance_2022, connell_performance_2021, nguyen_performability_2022, mendonca_performability_2020} and less than the number of requests fielded by typical servers. For example, the Wikimedia Foundation reports an average of 400 requests per second per machine across their servers \cite{wikimedia}, and Google Cloud allows a maximum of 1000 concurrent requests per instance \cite{google_cloud_cloud_2024}. 
This makes it a reasonable minimum rate of traffic for the attacker to require.  

Different numbers of servers and clients are used to demonstrate how the attack is affected by increasing the number of nodes in the network. 
The size of webpages available on the server ranges from 400kB to 4MB, following the typical page weights below the 75th percentile found on the internet as reported by the HTTP Archive \cite{httparchive}. The client requests webpages at random following a normal distribution with a mean file size of 2MB. 
Each set of experiments is run for 30 trials. For each trial, packets are sniffed from the network for the length of the observation window.

\begin{table}
\caption{Experimental Parameters and Their Values.}
\centering
\begin{adjustbox}{width=0.47\textwidth}    
    \begin{tabular}{|l |l |l|} \hline    
    \textbf{Variable} & \textbf{Notation} & \textbf{Range of values} \\ \hline 
    Attacker observation window & $O_{\textup{win}}$ & 3T – 10 hours \\ \hline 
    MTD interval (seconds) & $T$ & 30, 60, 120, 180, 240, 300, 600 \\ \hline 
    \# client requests/sec & $\lambda$ & 10, 5, 1, 1/2, 1/5, 1/15, 1/30, 2/T \\ \hline 
    \# servers & $n_{\textup{S}}$ & 1, 2, 3 \\ \hline 
    \# clients & $n_{\textup{c}}$ & 1, 2, 3 \\ \hline 
    Size of webpages & $W$ & 400kB – 4MB \\ \hline 
    Trials & $N$ & 30 \\ \hline
    \end{tabular}
\end{adjustbox}
\label{fig:parameters}
\end{table}

The exact set of parameters used for each experiment is specified in the results section. The data collected were the network packets sniffed by the attacker and the ground truth times when the MTD was triggered.

\section{Attacker's Approach} \label{Section:attack}
\begin{figure*}[t]
    \centering
    \includegraphics[width=0.95\textwidth]{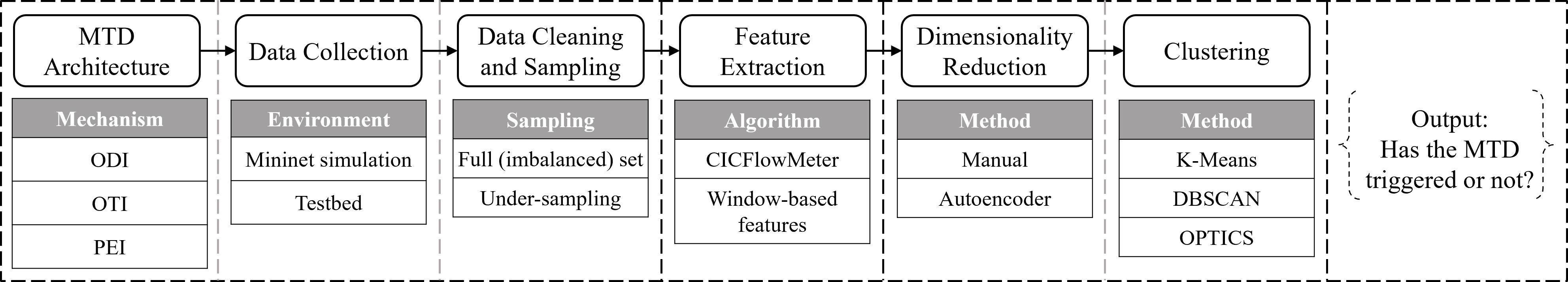}
    \caption{Proposed MTDSense approach for data analysis.}
    \label{fig:pipeline}
\end{figure*}

The complete MTDSense pipeline is shown in Figure  \ref{fig:pipeline}, and the details of specific steps are discussed below.

\subsection{Feature Set}
The features that are relevant to changes made by the MTD should capture changes in connections to a moving host in a network. These will appear across time and within flows rather than being distinguishable between individual packets. Therefore, both flow-based and window-based features were extracted from the raw packet data. The flow-based features are built on the CICFlowMeter-v4 feature extractor, which analyses flow traffic into 75 flow-level features \cite{sharafaldin_toward_2018}. Not all of these features are relevant to this problem, so those that were unchanged across all flows were removed at the outset. In addition to these, timing and delay features were added. These included the timing and delay of the TCP handshake and message body packets separately. The interarrival time between flows and entropy of IP addresses over time were also included, as these are expected to best directly capture the change to network components after an MTD trigger. An analysis of the importance of the relevance of the features is discussed in Section \ref{Section:eval:feat}.

Naturally, most of the time, the MTD is not being triggered. This means the volume of traffic captured directly after an MTD trigger is far less than the volume of traffic in between triggers. Therefore, there is a large class imbalance when distinguishing between the two cases, which was addressed with undersampling. 

\subsection{Clustering Techniques}

We assume that the attacker knows that the network is adopting an IP shuffling MTD but does not know the settings of the MTD and is not able to access the ground truth for the MTD trigger time or interval. As such, we must use unsupervised clustering algorithms to determine the MTD trigger time. To this end, we use primarily $k$-means clustering, which is a very commonly used method and continues to be used in some state-of-the-art clustering pipelines \cite{Ikotun}. Other clustering algorithms, such as  OPTICS, a generalization of the DBSCAN algorithm, which is density-based and can capture patterns of behaviors that $k$-means cannot, were also employed. However, after dimensionality reduction, differences in results were not pronounced. 

Given that there is over $80$ features and not all are relevant, we need a dimentionality reduction technique to obtain consistent results.    
The clustering techniques were combined with an autoencoder for dimensionality reduction. Firstly, this was done in two separate steps, where the autoencoder was trained and optimized for reconstruction loss, and the results were used to train the clustering algorithm. As an extension, the autoencoder and $k$-means algorithms were trained together with a combined loss function. This approach was introduced in \cite{xie_unsupervised_2016} and is commonly used as a benchmark in clustering literature.

The state-of-the-art in unsupervised clustering is in the image domain and uses generative adversarial networks \cite{pang_deep_2021}. However, the image data is significantly different from our data and has more complex spatial relationships. These models do not transfer well to our domain. Additionally, we found little difference before and after fine-tuning, which indicates we have already extracted the maximum information from the data. Postprocessing methods are discussed where they are relevant.

\subsection{Evaluation Metrics}

Since the data is being clustered unsupervised, alternative metrics that do not rely on labeling need to be used. Comparing labels will lead to scores that are not well defined on different permutations of the clustering and give results that are difficult to compare. Instead, the metrics need to measure how well the data is grouped compared to the ground truth. For this, we use the standard metrics often used in benchmarking clustering algorithms. This includes the adjusted Rand index (ARI) \cite{rand1971objective} and the clustering accuracy \cite{xie_unsupervised_2016} (which we report as `accuracy'). Note that ARI  ranges mostly between $0$ and $1$; however, it does not change linearly in comparison to accuracy. The ARI will be $0$ if the clustering is completely random and $1$ if the cluster and class set are identical in their pairwise membership. The value is bounded by $-0.5$, where negative values indicate the clustering is less in line with the truth labels than the expected random case.

\section{Evaluation} \label{Section:Eval}
In this section, we evaluate the attacker's ability to determine the timing of the MTD trigger with the following key questions. Firstly, how effectively can the attacker, with our proposed method, detect the MTD interval? Secondly, can modifying the method of triggering reduce the effectiveness of the detection? Thirdly, what conditions are needed for the attacker's approach to be effective?

\subsection{Determining That The MTD Has Been Triggered}

The results in this section use data with the On-Demand Installation update mechanism, which we consider to be the most natural for a network to implement because it most closely conforms to the asynchronous behavior of the internet. Section \ref{Section:eval:architectures} provides a comparison of attack results with other update mechanisms. The network configuration for this set of experiments comprised one server and one client, with data collected by the attacker and processed as described in Section \ref{Section:attack} to determine the network flows that follow an MTD trigger. In the case of the ODI update mechanism, this does not necessarily correspond to the absolute time the MTD triggered. Here, we report the MTD detection as successful for each item used in the ARI metric if the absolute time determined for the MTD trigger is within one second of the ground truth MTD trigger time, which is a small interval compared to the MTD interval $T$. We call this one second the `allowed detection delay' $s$. 

\begin{mdframed}
\noindent
The results demonstrate that the attacker can determine whether the MTD has been triggered with an ARI as high as $0.92$ and accuracy as high as $0.97$, depending on the experimental parameters.
\end{mdframed} The effect of network parameters on the fidelity of the results is explored below.

\begin{figure}
    \centering
    \includegraphics[width=0.48\textwidth]{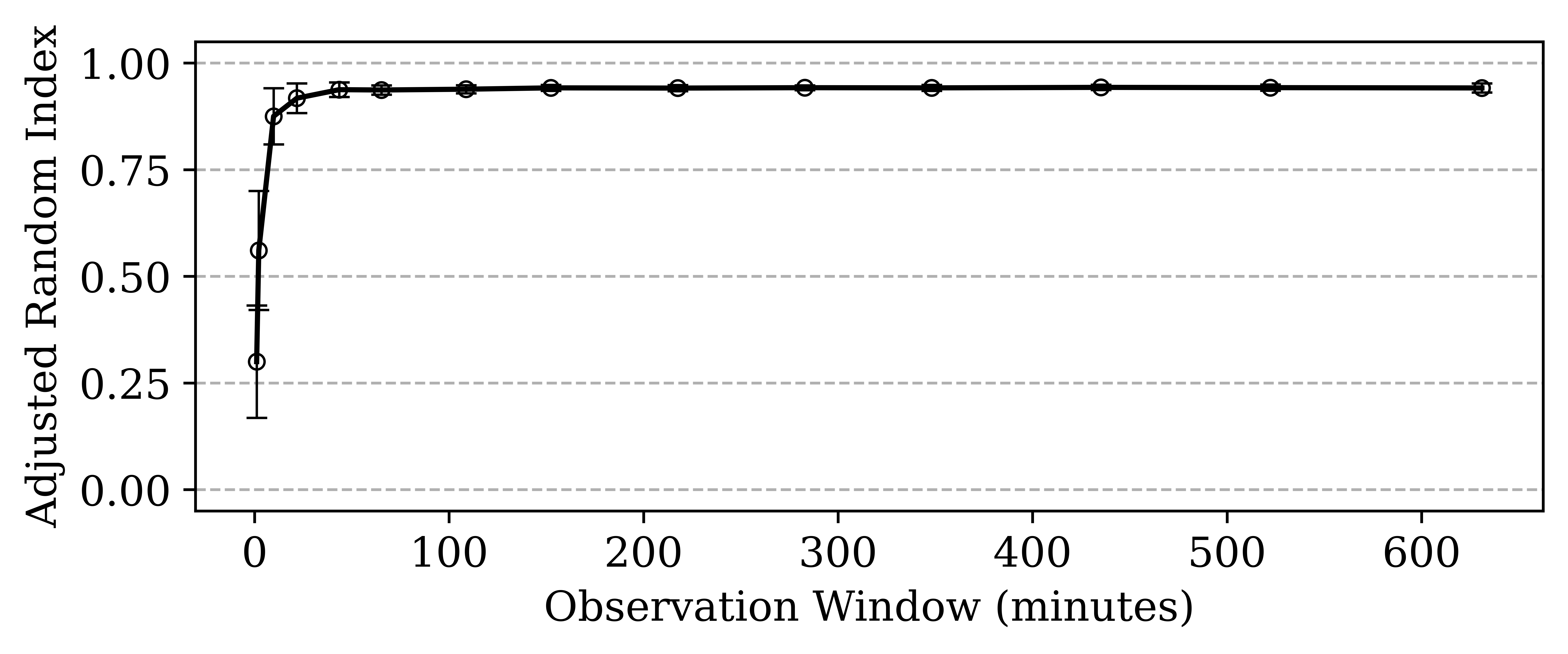}
    \caption{The effect of attacker observation window on ARI for $T=180s$.}
    \label{fig:result_obswindow}
\end{figure}

To determine the most appropriate observation window for the attacker to use, we tried possible values between one minute and 10 hours in an environment with an 180s MTD interval. In each case, the attacker collected data for the length of an observation window and clustered the data for that window. The results are shown in Figure \ref{fig:result_obswindow}, with each point being an average of 50 trials. Error bars show the $95\%$ confidence interval. Based on this information, an observation window of 2 hours was used for all other results reported going forward, but note that the attacker's information is fairly good with an ARI of $0.875$ and accuracy of $0.94$ for an observation window of 10 minutes, which is three times the MTD interval.\hfill
\begin{mdframed}
\noindent
This shows that an attacker with a window as small as $3T$ could still gain useful information with a few trials, and the effectiveness plateaus at a window of two hours. 
\end{mdframed}

\subsubsection{The Effect of the MTD Interval}
\begin{figure}
    \centering
    \includegraphics[width=0.48\textwidth]{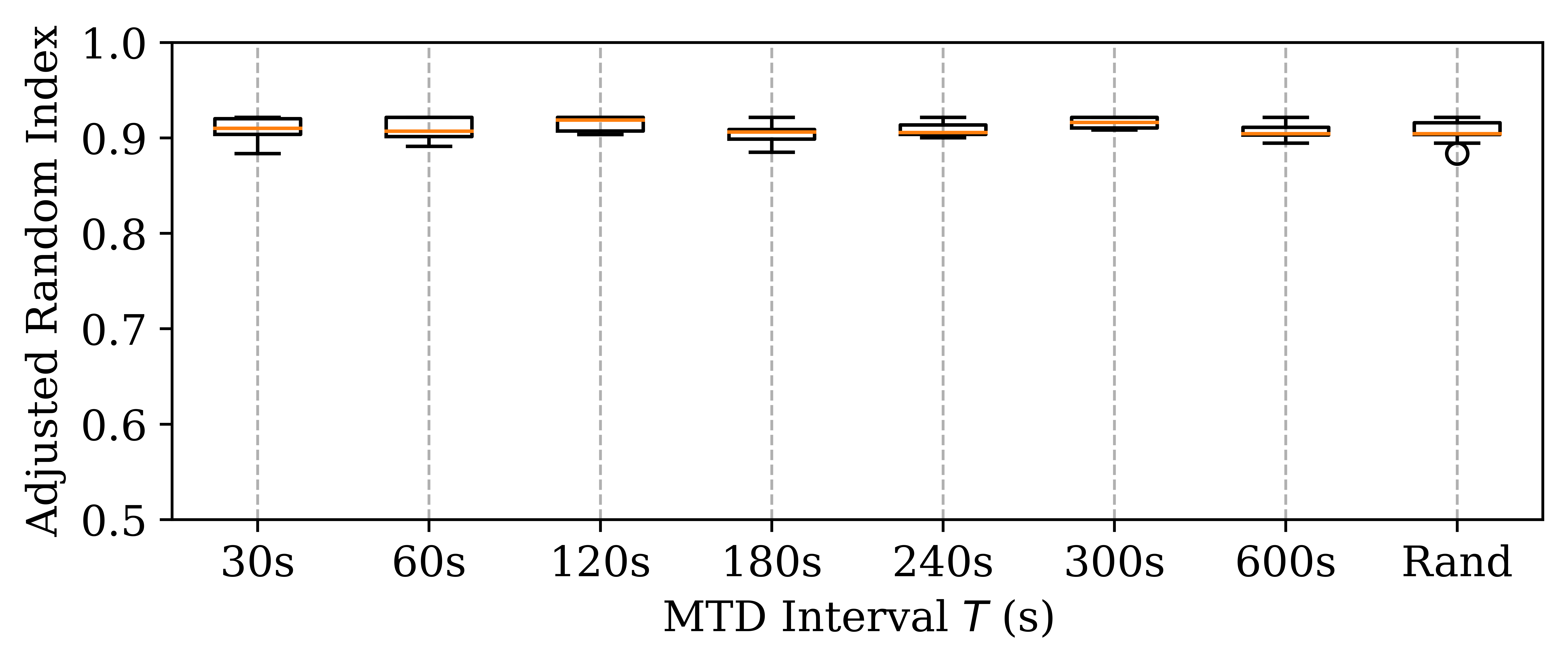}
    \caption{The effect of varying the MTD interval $T$ on the attacker's achieved ARI. The right-most result is for an MTD where the next trigger time is decided by sampling randomly from a distribution.}
    
    \label{fig:result_differentT}
\end{figure}

Figure \ref{fig:result_differentT} shows the ARI in detecting whether the MTD triggered or not for different MTD intervals. From the results we can see that the detection rate is independent of the MTD interval, the attacker can determine $T$ within this range with similar accuracy for all the values tested. Unlike the security benefit of MTD against traditional attacks that do not target the MTD mechanism directly, 
the accuracy of detecting that the MTD has triggered is not lowered by employing a lower MTD interval. As discussed in section \ref{Section:implementation:params}, we exclude very short MTD intervals which we deem to have too large a performance cost. Although very short MTD intervals may lower the detection accuracy, they would not be performant and so are unlikely to be employed in a network.  

\begin{mdframed}
We achieved a similar accuracy of detection when the MTD was triggered at a random interval. 
\end{mdframed}
For the final column of Figure \ref{fig:result_differentT} labeled `random', the MTD interval was sampled randomly from a normal distribution with a range of 15 to 300 seconds at each trigger. The symptoms of the MTD triggering in the network are completely independent of the cause of triggering and of $T$. Referring to the Algorithms  \ref{alg:od_install}-\ref{alg:pe_install}, the effects are due to the `newRule' installation steps and appear regardless of the condition that caused the MTD to trigger. For this reason, the method is still applicable to an MTD that is triggered by a random interval, and we predict it will also apply to an event-based MTD or any other triggering mechanism.

\subsubsection{The Effect of Client Traffic}
\begin{figure}
    \centering
    \begin{subfigure}{0.48\textwidth}
         \includegraphics[width=1\textwidth]{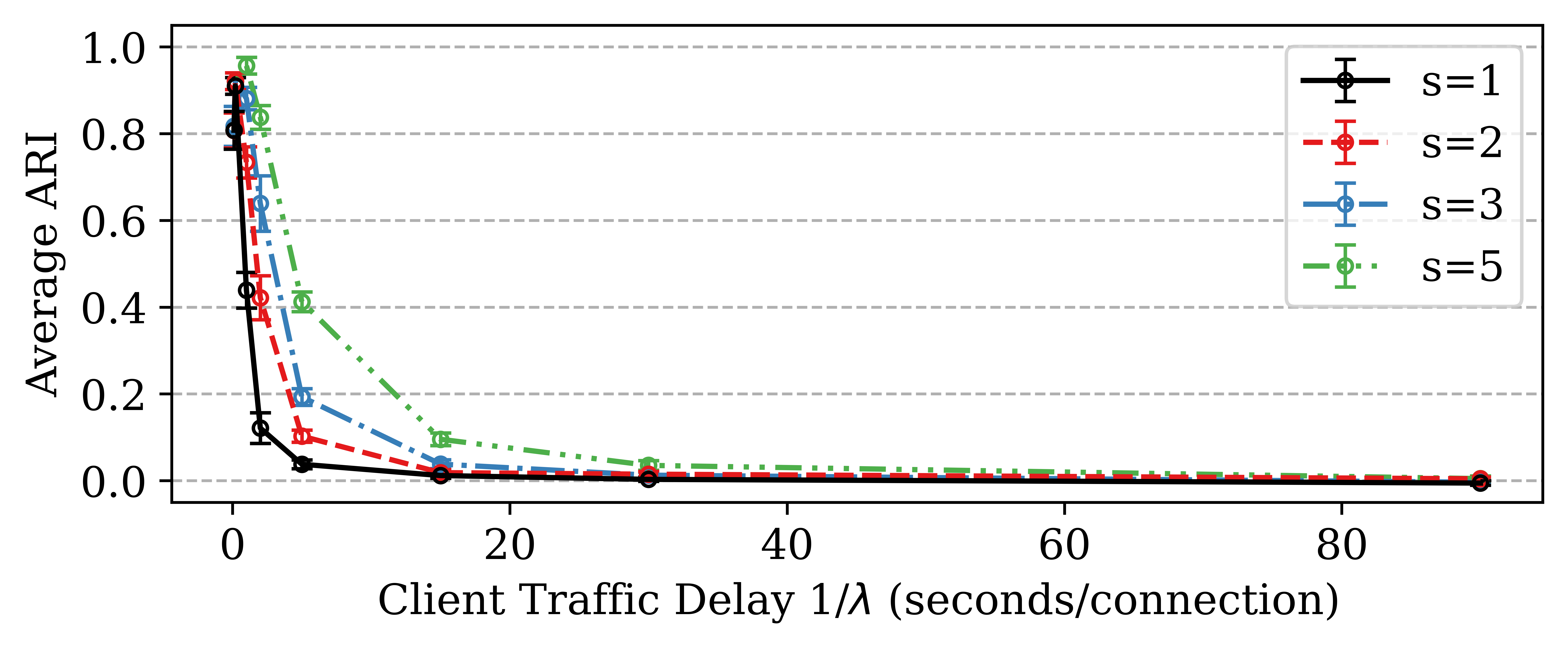}
         \caption{}
         \label{fig:result_differentLambda:nonlog}
    \end{subfigure}
    \vfill
    \begin{subfigure}{0.48\textwidth}
        \includegraphics[width=1\textwidth]{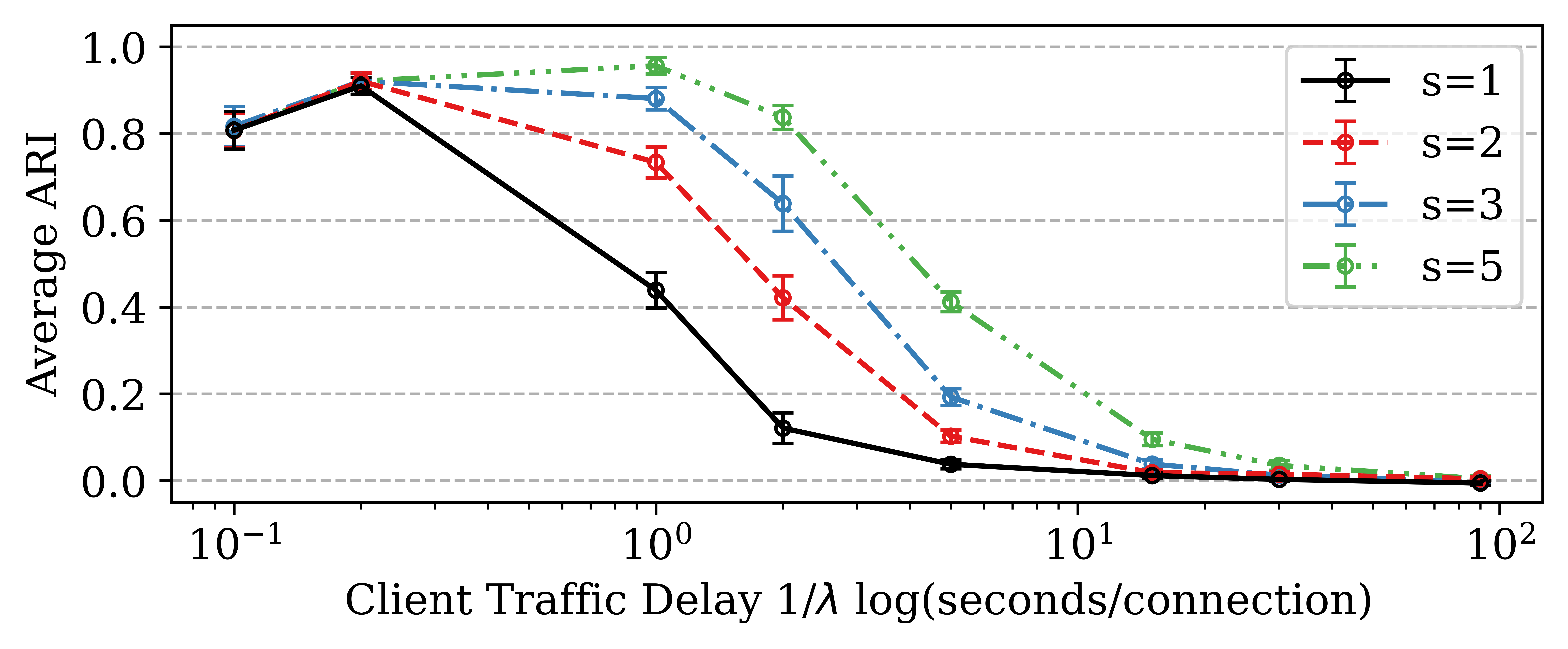}
        \caption{}
        \label{fig:result_differentLambda:log}
    \end{subfigure}
     
    \caption{\textbf{(a)} The effect of the rate of legitimate client traffic $\lambda$ on ARI. In \textbf{(b)}, the same data is shown over a log scale for clarity. Given over a range of allowed detection delays $s$.}
    \label{fig:result_differentLambda}
\end{figure}

Figure \ref{fig:result_differentLambda}  shows the effect of the average delay between client requests on the attacker's ability to determine the MTD has triggered. The rate of client traffic ranges from 10 connections per second to one or two connections per MTD interval, as discussed in Section \ref{Section:implementation:params}. Figure \ref{fig:result_differentLambda:nonlog}  shows the result over this full range, and \ref{fig:result_differentLambda:log} shows the results over a log axis so that the effect over the majority of the points is clearer.  

Focusing on the $s=1$ case, we can see the ARI is higher for lower delays until the lowest delay value, which resulted in less accurate detection than the second lowest value. This is due to the maximum throughput of our network and the delay between triggering the MTD within the network and updating the DNS server. Since the DNS update is not instantaneous, it is possible for a client connecting at the exact time between the update to receive the old vIP just before it becomes invalid and not be able to connect to it by the time they make their request to the network. The client-server connection will still be established since there is already a built-in timeout mechanism for requesting the address again and reconnecting.

However, this creates an additional performance degradation as the connection takes longer to establish. As the rate of requests increases, the chance of a request being made at the DNS update time increases, hence we see more of these events at the highest request rate. 
For the attacker determining that the MTD has triggered, this creates a different fingerprint in the traffic, and so the time is not detected accurately. 
\begin{mdframed}
\noindent
This means the ARI increases with increased rate of client requests until there are too many requests for the network to handle without some performance degradation, at which point the attacker's accuracy also decreases.
\end{mdframed}

The figure also shows the result for different allowed detection delays. The data must have ground truth values assigned to it. In the $s=1$ case, a positive label is assigned to a flow if it is within one second of the MTD trigger. We can see that in the $s=1$ case, the rate of client traffic must be higher than $5requests/s$ to give good detection, and it starts to degrade at $\lambda = 1s$. Results for the other $s$ values follow a similar pattern where $s=2$ degrades at $\lambda=2s$, and so on. Given the way the ground truth labels are assigned, this effect is less due to the attacker's ability to detect the MTD trigger and more to do with the ODI update mechanism.

When using the ODI update mechanism, the MTD rules are not installed in the OpenFlow switches until there is relevant network traffic. Due to this, we can consider the MTD trigger time to be transient, where there is an initial phase at the trigger time $T_i$ and a time when all the rules come into effect $t^c_i$, which is probabilistic. The time the rules are installed depends on when traffic arrives at a particular switch, so $t^c_i$ is dependent on the rate at which traffic arrives,  the number of paths through the network, and how those paths are utilized. The final rule can be installed at any time $t^c_i$ before the start of the next interval $T_{i+1}$.  
Because the ground truth times are the trigger time $T_i$ and the symptoms on the network depend on when the rules are installed $t^c_i$, the delay with which an attacker can determine that the MTD has triggered depends on the delay between when the MTD is triggered and when a client sends a packet. That is the the attacker's delay $\Delta t_A = \|\tilde{T}^A_i - T_i\|$ 
depends on the transient installation interval $\Delta t_I = \|t^c_i - T_i\|$. Since there is a causal relationship, the attacker's delay cannot be shorter than the transient period ($\Delta t_A \geq \Delta t_I$). Note from the figure that as the rate of client traffic increases, the duration of the transient period decreases, so the attacker can accurately determine the MTD trigger time with a lower detection delay. 

This delay due to the transient period is important because if the attacker wants to time their attacks to complete within an interval, they need to be able to accurately predict the duration when the current set of IP addresses is valid, and this duration is between the ground truth times ${T_i}$ and $T_{i+1}$, with no fixed relationship to $t^c_i$. In the ODI case, the attacker can always tell whether the MTD has been triggered since some previous time, even when there is no traffic at the time of the MTD trigger. This includes the cases where there is only one connection or less in each MTD interval. The attacker can determine their confidence of when the MTD is triggered based on the rate of client traffic.  This also means that in the ODI scheme, the attacker can calculate the MTD interval $T$ accurately based on the timings of $t_{ci}$, even when there is not enough traffic to accurately calculate the time $t_{Ti}$.  
This is addressed in Section \ref{Section:eval:extractingT}. The effect of the rate of traffic in the OTI and PEI MTDs is explored in the next section. 

\subsubsection{Extracting The MTD Interval} \label{Section:eval:extractingT}
Given that the attacker knows when the MTD has triggered, they can restart their attack to maximize the possible attack window. However, if the attacker knows the MTD interval, they can further plan their attacks and limit them to those that will have the greatest chance of succeeding within the interval. 
To this end, we try to estimate the MTD interval $T$ and report the accuracy with which it can be detected, given the accuracy of detecting the MTD trigger. The results are shown in Table \ref{table:results_extractingT}. Note that all values are predicted with an error of less than one percent.

\begin{table}
    \caption{The MTD Interval $T$ As Predicted by the Attacker Based on Their Clustering. }
    \centering
    \begin{tabularx}{0.47\textwidth}{|C|C|C|C|C|} \hline 
         &  \multicolumn{2}{|c|}{\textbf{Mininet Data}}&  \multicolumn{2}{|c|}{\textbf{Testbed Data}}\\ \hline 
         \textbf{Actual \newline Interval \newline (seconds)}&  \textbf{Predicted Interval (seconds)}&  \textbf{Error Rate}&  \textbf{Predicted Interval (seconds)}& \textbf{Error Rate}\\ \hline 
         60&  59.73&  0.45\%&  60.39& 0.65\%\\ \hline 
         180  & 179.58   & 0.23\%   & 180.84   & 0.47\%   \\ \hline
         300&  300.32&  0.11\%&  301.58 &  0.52\\ \hline 
    \end{tabularx}
    \label{table:results_extractingT}
\end{table}

\subsubsection{Feature Importance}\label{Section:eval:feat}
\begin{figure}
    \centering
        \includegraphics[width=0.45\textwidth]{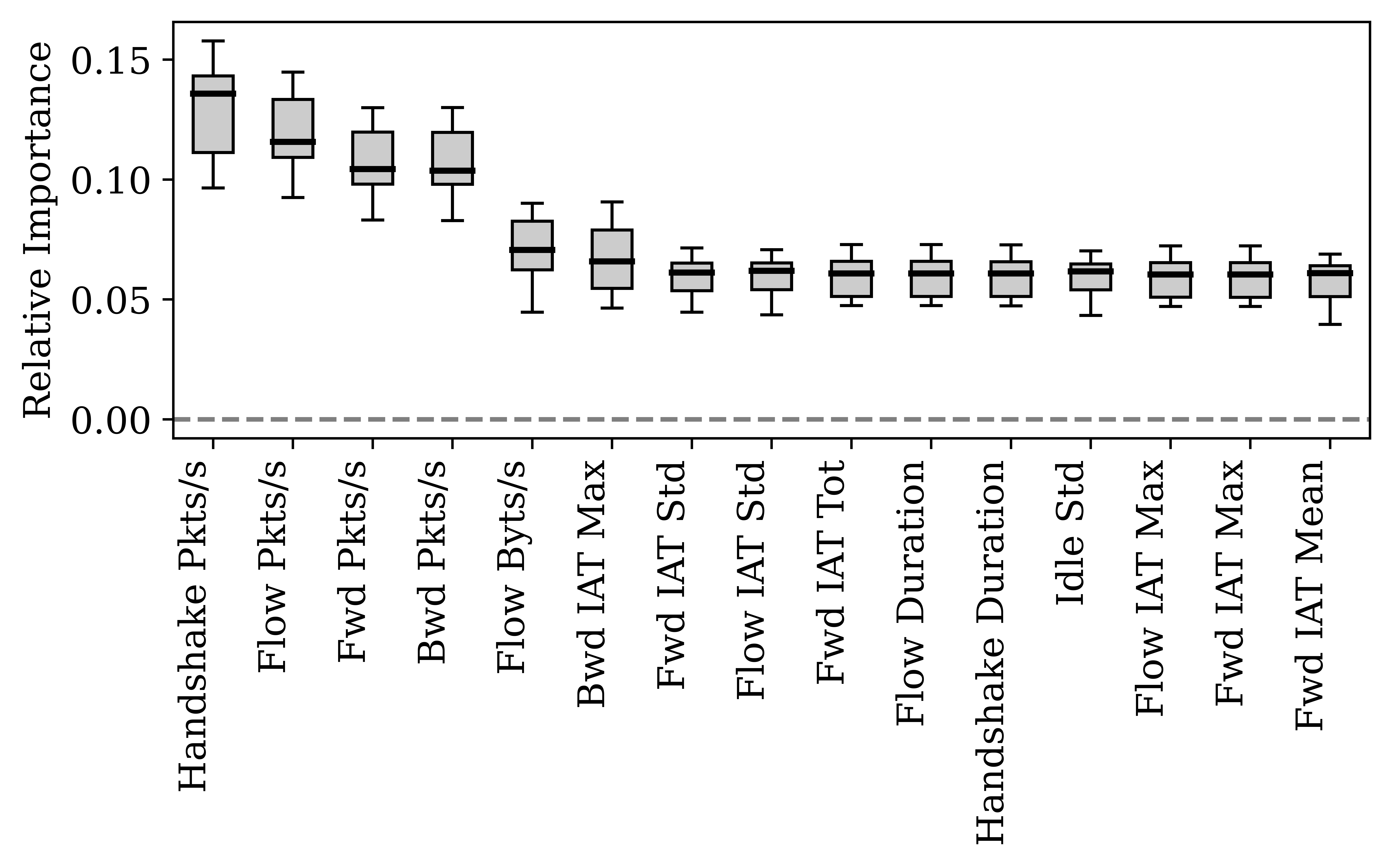}
    \caption{Ranking of feature importance to clustering.}
    \label{fig:feat_importance}
\end{figure}

To rank the importance of the features to the clustering result, we use the approach proposed by  Ismaili \textit{et al.} \cite{ismaili_supervised_2014}, where a random forest classifier is trained with the clustered labels as the training labels, and the feature importance is extracted from the classifier.  The results for the 15 most important features are shown in Figure \ref{fig:feat_importance}.

We note that all of the most important features were related to the timing features of the traffic, specifically to the rate of connections and the difference in rate between the TCP handshake and the rest of the connection. This indicates the MTD trigger is detectable due to the delay it creates in the network traffic when rules are installed. This suggests that detectability can be addressed by making the timing of the MTD more closely resemble normal network timings, which leads to the idea of alternative update schemes that aim to reduce the delay associated with the MTD trigger. 

\subsection{Effect of MTD Update Mechanism} \label{Section:eval:architectures}
\begin{figure}
    \centering

    \begin{subfigure}{0.48\textwidth}
         \includegraphics[width=1\textwidth]{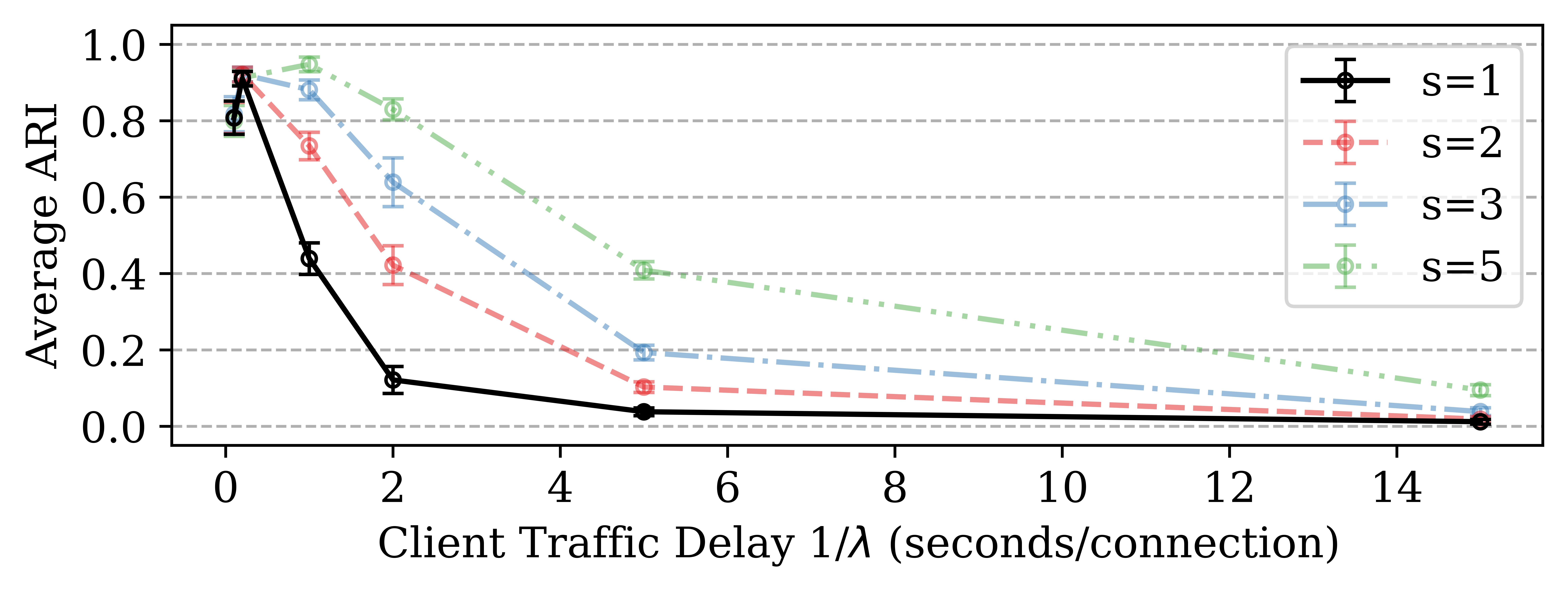}
         \caption{}
         \label{fig:result_OTI:odi}
    \end{subfigure}
    \vfill
    \begin{subfigure}{0.48\textwidth}
        \includegraphics[width=1\textwidth]{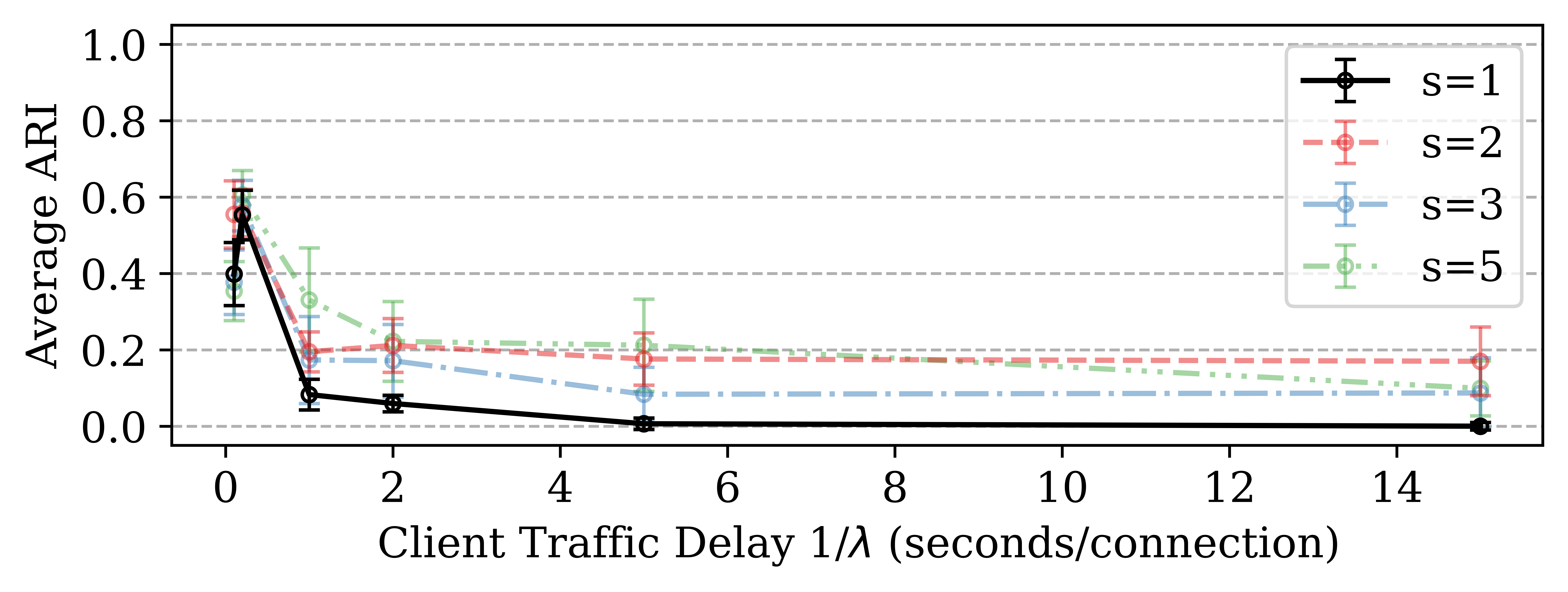}
        \caption{}
        \label{fig:result_OTI:oti}
    \end{subfigure}
    
    \caption{ARI for clustering the MTD triggers with \textbf{(a)} the ODI update mechanism, and \textbf{(b)} the OTI update mechanism, $T=180s$.}
    \label{fig:result_OTI}
\end{figure}

We explored whether the other proposed update mechanisms affect the attacker's ability to determine that the MTD has been triggered. 
Experiments were run using the OTI and PEI update mechanisms with an MTD interval of 180 seconds and for a range of client request frequencies $\lambda$. Each data point is the result of 30 two-hour trials, and the results can be compared to those in Figure \ref{fig:result_differentLambda}. The results for the OTI mechanism are shown in Figure \ref{fig:result_OTI:oti}.  For convenience in comparison, the results for the  ODI update mechanism over a smaller range are included as Figure \ref{fig:result_OTI:odi}. We can see the attacker achieved a similar accuracy for both schemes when there was more client traffic ($\lambda$ was high). As the amount of client traffic drops, the accuracy drops somewhat faster with OTI. In OTI, the rules are installed in the switch at the determined MTD time, which takes some slightly variable amount of time $\eta$. If the delay between the installation time and the arrival of client traffic is more than $\eta$, there will be no symptom of the MTD trigger in the network.  However, at high $\lambda$ where traffic is more likely to arrive within $\eta$, the attacker is still effective.  Given that in a network, the utilization of each server is maximized as much as possible, this change may not be sufficient. 

Also note that in the OTI case, there is no longer an effect due to $s$. At lower delays all $s$ achieve similar accuracies, and at higher delays, all $s$ have large overlapping error bars indicating the results are noise. This is because there the detection here is tied to $T_i$ and not $t^c_i$, and therefore detection is not possible at lower rates of traffic.  

\begin{figure}[t]
    \centering
    \includegraphics[width=0.5\textwidth]{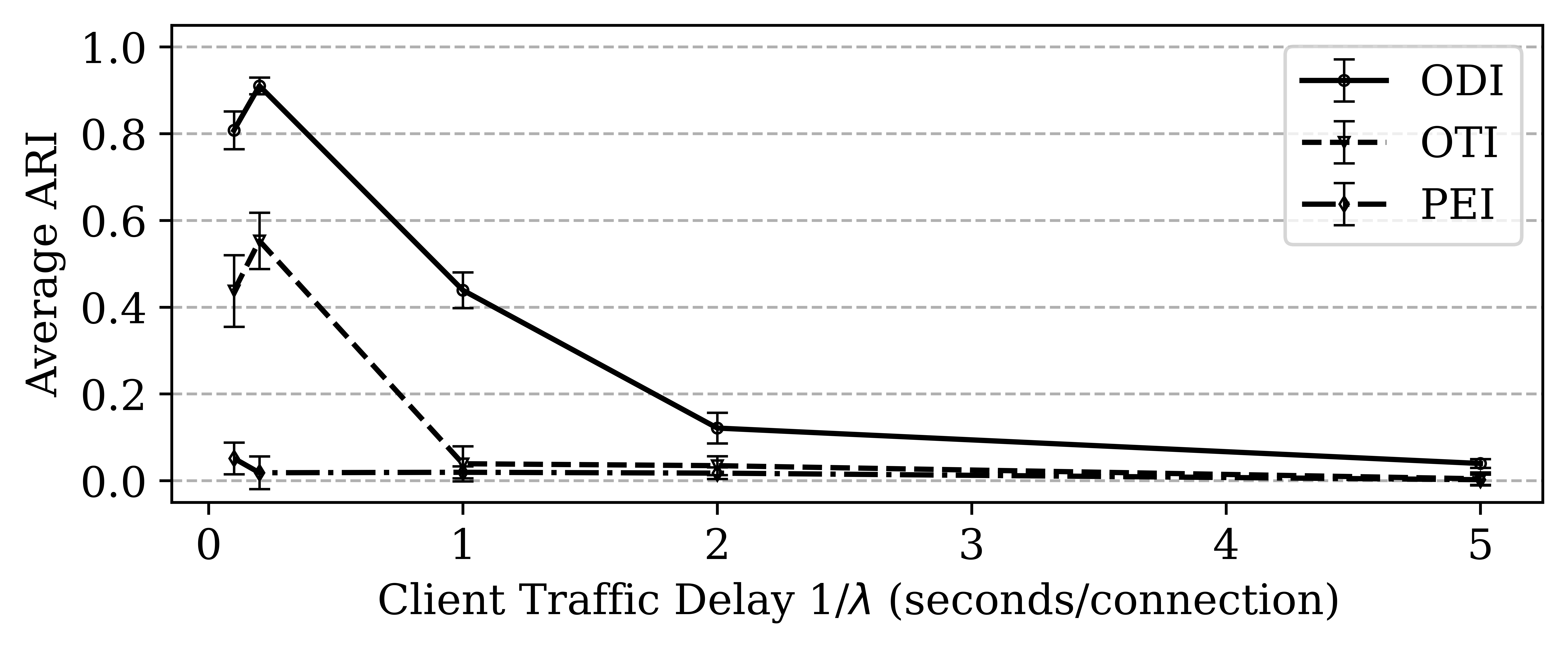}
    
    \caption{Comparison of ARI for clustering the MTD triggers between all update mechansims,  $T=180s$.}
    \label{fig:result_PEI}
\end{figure}

Figure \ref{fig:result_PEI} shows the results for the PEI mechanism with the ODI and OTI cases overlaid for comparison. The attacker has almost no ability to determine the MTD has triggered here, even at higher $\lambda$.  This indicates that the PEI mechanism, making updates before they are needed, is an effective way to reduce the timing difference before and after the MTD trigger and thus remove the attacker's ability to detect the trigger. We note the detection here was variable, with some triggers being detected correctly while others were missed. In other words, there were more false negatives, but true positives were still present, while false positives were very low and true negatives high.  The attacker is still able to detect some of the MTD triggers at high $\lambda$. 

The theory for PEI aims to hide the MTD trigger entirely by performing all necessary changes preemptively, so there should be no symptoms of the trigger in the network. The results show some symptoms are still present, which we traced back to the timing of the network. While the MTD trigger is no longer responsive, there is still a delay in updating cached DNS records. Additionally, because updates are asynchronous, it is not always possible to synchronize all switches to update exactly at the same time. Given that the client requests can be so frequent, this means in some triggers, at least one switch still made a request to the controller.  

\begin{mdframed}
\noindent
The results show that we are able to minimize the range of amount of traffic the attacker's approach is effective for with OTI  and significantly reduce the detection with PEI. 
\end{mdframed}
In the PEI case, some symptoms are still present in the network, but the attacker's detection is significantly dampened. When considering whether PEI can be deployed as a defense mechanism against this attack, we have to consider that there is a large performance cost associated with the non-asynchronous update mechanism (OTI and PEI). These performance overheads are detailed in Section \ref{alg:pe_install}. Given that there is already a performance cost associated with implementing OTI MTDs, which has to be contended with, this may not be desirable in a deployed network.

\subsection{Effect of Network Size}
\begin{figure}
    \centering
    \includegraphics[width=0.48\textwidth]{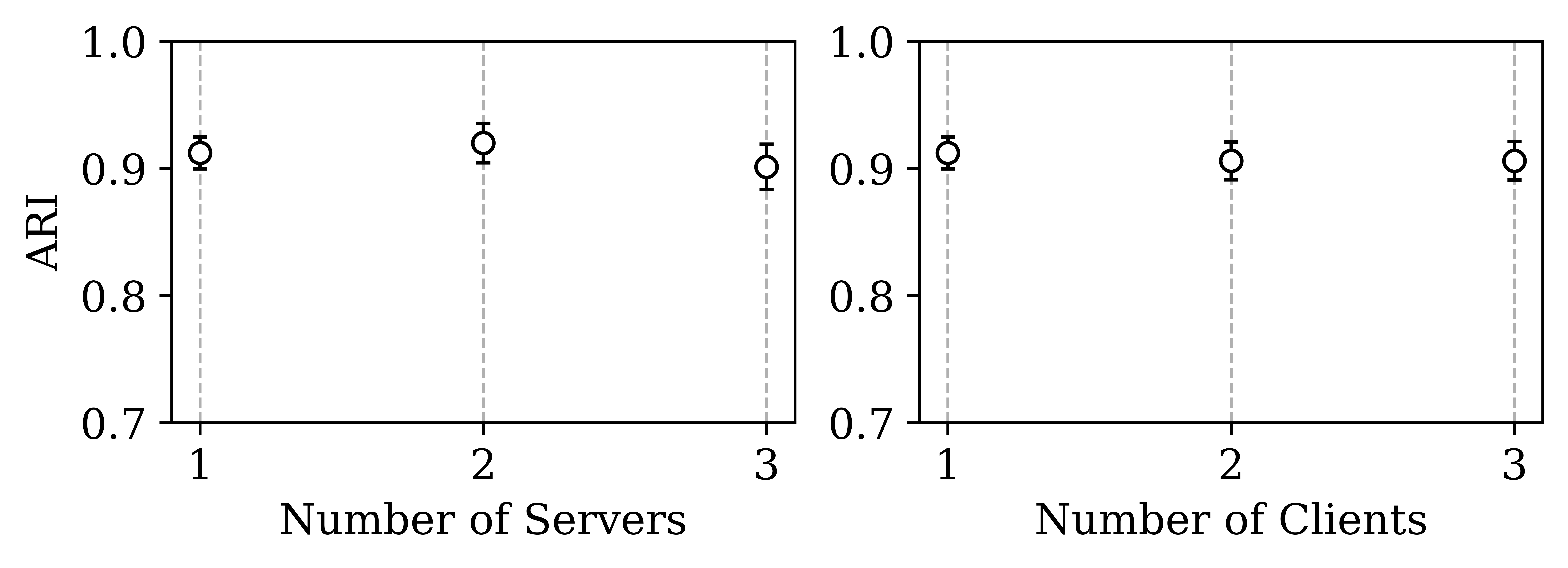}    
    \caption{The effect of the number of clients and servers on ARI when the total traffic volume across all hosts is kept constant. }
   
    \label{fig:results_clients}
\end{figure}

In this section, we first show that the total volume of client traffic, which makes the attack possible, need not necessarily come from one client or to one server. Under this model, we assume the attacker has compromised a series of devices that allow them to sniff the traffic from multiple clients to the network. The total rate of traffic from all clients $\lambda_{tot}$ to all servers is kept the same but is generated by one, two, or three clients. The network utilizes an 180s ODI MTD. The results are shown in Figure \ref{fig:results_clients}.

The detection pipeline has to be expanded by one step in this scenario because, in the ODI scheme, flow rules are installed for each client and server pair at each switch. This means if we have $n_{\textup{c}}$ clients and $n_{\textup{S}}$ servers,  after each MTD interval, $n_{\textup{c}} \times n_{\textup{S}}$ new rules are installed. Thus, the attacker will detect  $n_{\textup{c}} \times n_{\textup{S}}$ MTD triggers that are all very close in time. When determining $T_i$, the attacker now groups all triggers it detects within two seconds as one MTD trigger, with the assumption that this is a far higher trigger frequency than would be expected of a network in practice. 

\begin{mdframed}
\noindent
The results show similar effectiveness across one client and server to multiple clients and servers. A minimum total rate of traffic from all clients to all servers is required for the attacker to be effective. However, allowing the attacker to collect this traffic across many clients relaxes the assumptions.  
\end{mdframed}

\subsection{Cross Dataset Model Performance}
The experimental results so far show that the attacker's method for determining the MTD interval is effective across a range of MTD intervals in both a simulated and \textit{testbed} environment. These results are achieved with 2 hours of training data. However, a key question to the practicality of this method is what the attacker can do if they do not have access to the network to collect enough training data. Additionally, what if the MTD is responsive, and so its settings change during or after the training interval? In this section, we show that trigger detection is effective across datasets. 

Table \ref{table:results_crossdataset} shows the performance of the attacker's model with different training and testing environments using the ODI update mechanism. 
The data with the configuration in the row is used to train the autoencoder and clustered. Using the same autoencoder weights and cluster centers, the data from the configuration in the column is clustered, and the ARI is calculated. As an example, the value in row 1 column 3 shows the result of training the model with data from a simulated network with 180s MTD and testing it against the testbed with 60s MTD.  Across the diagonal is the performance of the models trained and tested from data in the same environment (though the training and test sets are separate). We can see from this table that the performance of the model does not drop significantly through this change. This is reasonable as the symptoms in the network traffic are not changed by changing the settings of the MTD, and as discussed in Section \ref{Section:eval:feat}, the features that contribute most to the clustering are timing and rate-based features, which when normalized create the same pattern in the two environments. 

\begin{mdframed}
\noindent
This indicates that the attacker's approach can still be successful if the settings of the MTD change and if the attacker cannot collect enough data from one set system. 
\end{mdframed}
So if one client cannot be completely monitored, the attacker can still be successful. Further, if the attacker still does not have enough access to the system but does know the overall design of the MTD, they can transfer a model from a different environment to the target environment. This may be applicable, for example, if the target is using a configurable commercial MTD. In such a case, the importance of ensuring the MTD is robust to this type of approach is very evident. 

\begin{table}[]
    \caption{Cross-dataset Performance of the Attacker Models. Results are given for both the simulation and testbed over different MTD intervals.}
\centering
\begin{tabular}{|cc|cccc|}
\hline
                                                                &                                                              & \multicolumn{4}{c|}{Testing Dataset }                                                                                                                                                                                                                                                                   \\ \cline{3-6} 
                                                                &                                                              & \multicolumn{1}{c|}{\begin{tabular}[c]{@{}c@{}}Simulation \\ T=60s\end{tabular}} & \multicolumn{1}{c|}{\begin{tabular}[c]{@{}c@{}}Testbed\\  T=60s\end{tabular}} & \multicolumn{1}{c|}{\begin{tabular}[c]{@{}c@{}}Simulation \\ T=180s\end{tabular}} & \begin{tabular}[c]{@{}c@{}}Testbed\\ T=180s\end{tabular} \\ \hline
\multicolumn{1}{|c|}{}                                          & \begin{tabular}[c]{@{}c@{}}Simulation \\ T=60s\end{tabular}  & \multicolumn{1}{c|}{\cellcolor[HTML]{E1E1E1}0.917}                               & \multicolumn{1}{c|}{0.92}                                                    & \multicolumn{1}{c|}{0.922}                                                        & 0.914                                                   \\ \hhline{~|-|-|-|-|-|} 
\multicolumn{1}{|c|}{}                                          & \begin{tabular}[c]{@{}c@{}}Testbed \\ T=60s\end{tabular}      & \multicolumn{1}{c|}{0.918}                                                       & \multicolumn{1}{c|}{\cellcolor[HTML]{E1E1E1}0.916}                           & \multicolumn{1}{c|}{0.911}                                                        & 0.892                                                   \\ \hhline{~|-|-|-|-|-|}
\multicolumn{1}{|c|}{}                                          & \begin{tabular}[c]{@{}c@{}}Simulation \\ T=180s\end{tabular} & \multicolumn{1}{c|}{0.921}                                                       & \multicolumn{1}{c|}{0.897}                                                   & \multicolumn{1}{c|}{\cellcolor[HTML]{E1E1E1}0.912}                                & 0.918                                                   \\ \hhline{~|-|-|-|-|-|} 
\multicolumn{1}{|c|}{\multirow{-4}{*}[6ex]{\rotatebox[origin=c]{90}{Training Dataset}}} & \begin{tabular}[c]{@{}c@{}}Testbed   \\ T=180s\end{tabular}   & \multicolumn{1}{c|}{0.899}                                                       & \multicolumn{1}{c|}{0.906}                                                   & \multicolumn{1}{c|}{0.913}                                                        & \cellcolor[HTML]{E1E1E1}0.923                           \\ \hline
\end{tabular}

    \label{table:results_crossdataset}
\end{table}

\section{Related Work} 
\label{Section:related}

Most literature on evaluating the effectiveness of MTDs has focused on assessing them against simple automated attackers that usually target traditional networks. Common types of attacks considered in network MTDs include scanning \cite{achleitner_deceiving_2017, jafarian_spatio-temporal_2014, luo_rpah_2015} and DDoS attacks ~\cite{zhou_cost-effective_2019, steinberger_ddos_2018,meier_nethide_2018, aydeger_mitigating_2016}. While these works demonstrate the effectiveness of MTD techniques, the attacks considered are not tailored to an environment with MTD and assume the attacker has no knowledge that an MTD is being used. 

Work considering an intelligent attacker that takes the MTD into consideration is very sparse. Jafarian \textit{et al.} evaluated a combined MTD and deception technique using six skilled human attackers who were able to identify hosts using their fingerprints (open ports and running services), making the MTD alone much less effective~\cite{jafarian_multi-dimensional_2016}. However, they used a long MTD interval of 15 minutes and did not randomize port numbers. Additionally, this requires the full attention of skilled human operators, which is costly for an attacker.     Moghaddam \textit{et al.} showed that if the attacker is assumed to know the MTD interval, they can plan their attacks to significantly increase the chances of success across all phases of a cyber attack \cite{moghaddam_practical_2022}.  However, they assumed the exact MTD interval is provided to the attacker, which is a strong assumption that is not realistic in typical attack environments.

There are also some works targeting MTD techniques directly in the domain of MAC address randomization MTDs in mobile phone devices. Vanhoef \textit{et al.}  uniquely identified and tracked mobile devices employing MAC address randomization MTD by clustering Wi-Fi probe requests \cite{vanhoef_why_2016}. This work demonstrated the ability to identify devices despite address mutation MTDs. However, it required that the protocol leak data by using non-standard headers and failing to change sequence numbers and scrambler seeds after mutation.  In a similar study,  Matte \textit{et al.} identified mobile devices from Wi-Fi probes using only the timing of the probes, without the need to rely on leaked information \cite{matte_defeating_2016}. These works circumvent the MTD instead of fingerprinting it but are notable in their use of clustering. To the best of our knowledge, no prior work has used machine learning or any other method to detect the MTD trigger and MTD interval in a comprehensive manner.

\section{Limitations}\label{limitations}
The main limitations of this work and possible areas for future research are as follows.

\textbf{MTD dimension}.  MTD has three dimensions: 1) when to move, 2) what to move, and 3) how to move~\cite{cho_toward_2020}. In this work, MTDSense was able to detect information about the first dimension: when to move. In our SDN setup, we have implemented other MTD techniques, such as virtual port shuffling and web application diversity, which will allow us to collect datasets involving the `what' and `how' aspects of MTD. In future work, we plan to extend the attacker's scope to determine the other two dimensions of MTD using the same techniques.

\textbf{Extension on client(s) traffic.} The effectiveness of MTDSense is reliant on active clients in the network creating a minimum amount of traffic. This gives the attacker flexibility in what components are compromised aids in evading detection, however is ineffective if there is not enough client traffic. It is possible that some active scanning by the attacker can be incorporated in environments where client traffic is sparse. Moreover, our experiments are conducted with very small numbers of clients and servers. These should be scaled to larger environments, which have additional complexities.

\textbf{Evaluation of other defense against MTDSense.} We explore alternative mechanisms for implementing MTD as a possible defense. However, these do not completely hide the operation of the MTD and have associated performance costs. Other defenses are possible and need to be explored.

\section{Conclusions} 
\label{Section:conclusion}

This work has presented a novel attacker approach named \mbox{MTDSense}. We have shown that by clustering network flows eavesdropped from an SDN network, the attacker can determine when the MTD is triggered and calculate the MTD interval. To our knowledge, this is the first work to estimate the MTD trigger interval using an AI-based approach. The experimental results have shown this attacker approach is effective for a range of traffic volumes and network sizes, and the learning can be transferable between networks with differing configurations. Additionally, we have proposed two new MTD update mechanisms and evaluated them against MTDSense, showing that they are effective at low traffic volumes, although we predict that they have considerable performance and privacy costs. The attacker's capability to determine the MTD interval significantly reduces the effectiveness of the defense and raises significant questions about the benefits of MTD. 

\balance
\bibliographystyle{IEEEtran}
\bibliography{main}

\end{document}